# The Insulin Superfamily of Growth-Promoting Proteins

Razvan Tudor Radulescu

*Molecular Concepts Research (MCR), Munich, Germany*

E-mail: ratura@gmx.net

**ABSTRACT**

Recently, structural analysis of the human transferrin and growth hormone (GH) amino acid sequences has unravelled that they harbor a motif identical to a pattern found in viral oncoproteins known to bind the primarily nuclear tumor suppressor retinoblastoma protein (RB). Since related signatures had previously been identified also in insulin and the two insulin-like growth factors (IGFs), the aim of the current study has been to investigate whether further hints substantiating these reported homologies can be found *in silico*. Here, additional similarities are presented supporting the notion of an insulin superfamily of growth-promoting proteins with dual localization in the extracellular environment and the intracellular space, particularly in the nucleus, as well as characterized by a tropism for RB.

Increasing evidence suggests that major growth-regulatory proteins employ similar amino acid information codes for directing cells towards quiescence, differentiation or proliferation. In this context, I have initially deciphered a potential for insulin, IGF-1 and IGF-2 to bind to the key cell cycle regulator and tumor suppressor RB (1), thus paralleling similar structure-function relationships in certain viral oncoproteins (2). I further predicted that, as a result of this physical interaction anticipated to take place primarily in the cell nucleus, RB would be inactivated and such mechanism may constitute an important growth-stimulatory event in both embryogenesis and oncogenesis (1). Subsequently, this presumed complex formation between insulin and RB was validated experimentally (3-6). So was the association between any of the two IGFs and an RB fragment (7) that had been proposed to represent the RB binding site for insulin, IGF-1 or IGF-2 (1).

Recently, I have identified the same RB-binding motifs also in transferrin (8) and in GH (9,10). Remarkably, GH contains a putative RB-binding signature that is homologous to two sequences present in HTLV-1 Tax, a viral oncoprotein recently shown to physically interact with RB (11). This feature may be particularly interesting given the nuclear translocation (12) and oncogenic properties (13,14) of GH.

Here, I reveal that the homologies between RB-binding viral oncoproteins and essential cellular factors with growth-promoting activity concern several motifs (Fig. 1), thus implying the existence of a superfamily of host proteins hitherto unknown.

| | | |
|---|---|---|
| $^{46}$**L** Y D **L**$^{49}$ ... | $^{122}$**L** T **C** H **E**$^{126}$ | Ad5 E1A |
| $^{12}$**M** L D **L**$^{15}$ ... | $^{22}$**L** Y **C** Y **E**$^{26}$ | HPV-16 E7 |
| $^{16}$**L** L G **L**$^{19}$ ... | $^{103}$**L** F **C** S **E**$^{107}$ | SV40 large T |
| $^{205}$**L** I I **L**$^{208}$ ... | $^{307}$**L** L F N **E**$^{311}$ | HTLV-1 Tax |
| $^{305}$**L** H L **L**$^{308}$ ... | $^{319}$**L** L F N **E**$^{323}$ | HTLV-1 Tax |
| $^{20}$**L** H Q **L**$^{23}$ ... | $^{52}$**L** **C** F S **E**$^{56}$ | human GH |
| $^{225}$**L** L **C** **L**$^{228}$ ... | $^{353}$**L** K **C** D **E**$^{357}$ | human transferrin |
| $^{37}$**L** **C** L **L**$^{40}$ ... | $^{64}$**F** V **C** G **D**$^{68}$ | human IGF-1 precursor |
| $^{13}$**L** T F **L**$^{16}$ ... | $^{43}$**F** V **C** G **D**$^{47}$ | human IGF-2 precursor |
| $^{13}$**L** L A **L**$^{16}$ ... | $^{41}$**L** V **C** G **E**$^{45}$ | human preproinsulin |
| $^{13}$**L** Y Q **L**$^{16}$ | | human insulin A-chain |
| | $^{17}$**L** V **C** G **E**$^{21}$ | human insulin B-chain |

Fig. 1   RB-binding amino acid motifs (LXXL and LXCXE as well as patterns related thereto whereby X stands for any amino acid) in viral oncoproteins and cellular growth factors. Crucial residues are highlighted in bold letters.